\documentclass[]{spie}  
\usepackage[]{graphicx}

\newcommand{\etal}{et al.}

\newcommand{\eg}{e.g.}

\newcommand{\micron}{\mbox{$\mu{\rm m}$}}

\newcommand{\Mjup}{\hbox{M$_{\rm Jup}$}}

\newcommand{\arcsec}{\mbox{$^{\prime \prime}$}}
\newcommand{\arcmin}{\mbox{$^{\prime}$}}

\newcommand{\HST}{{\sl HST}}
\def\lesssim{\mathrel{\hbox{\rlap{\hbox{%
 \lower4pt\hbox{$\sim$}}}\hbox{$<$}}}}
\def\gtrsim{\mathrel{\hbox{\rlap{\hbox{%
 \lower4pt\hbox{$\sim$}}}\hbox{$>$}}}}

\newcommand\aj{AJ}%
\newcommand\apj{ApJ}%
\newcommand\apjl{ApJ}%
\newcommand\aap{A\&A}%
\newcommand\aaps{A\&AS}%
\newcommand\pasp{PASP}%
\newcommand\nat{Nature}%

\title{Astronomical Science with Laser Guide Star Adaptive Optics:\\ A
  Brief Review, a Current Snapshot, and a Bright Future}

\author{Michael C. Liu\supit{a}
\skiplinehalf
\supit{a} Institute for Astronomy, University of Hawai`i, 2680
  Woodlawn Drive, Honolulu, HI 96822; USA 
}

\authorinfo{Email: mliu@ifa.hawaii.edu}

\begin{document} 
  \maketitle 

\begin{abstract}
We briefly discuss the past, present, and future state of astronomical
science with laser guide star adaptive optics (LGS AO).  We present a
tabulation of refereed science papers from LGS AO, amounting to a total
of 23 publications as of May 2006.  The first decade of LGS AO science
(1995--2004) was marked by modest science productivity
($\approx$1~paper/year), as LGS systems were being implemented and
commissioned.
%
%
The last two years have seen explosive science growth
($\approx$1~paper/month), largely due to the new LGS system on the
Keck~II 10-meter telescope, and point to an exciting new era for high
angular resolution science.  To illustrate the achievable on-sky
performance, we present an extensive collection of Keck LGS performance
measurements from the first year of our brown dwarf near-IR imaging
survey.  We summarize the current strengths and weaknesses of LGS
compared to {\sl Hubble Space Telescope}, offer a list of desired
improvements, and look forward to a bright future for LGS given its
wide-scale implementation on large ground-based telescopes.
\end{abstract}



\section{INTRODUCTION}
\label{sect:intro}  

Astronomers have envisioned using laser guide star adaptive optics (LGS
AO) to achieve diffraction-limited observations from ground-based
telescopes for over two decades\cite{1985A&A...152L..29F,
  1987Natur.328..229T, 1994OSAJ...11..263H}.  (See
Refs.~\citenum{2004aoa..book.....R} and~\citenum{1998aoat.conf.....H} for a
historical review of LGS AO development.)  The realization of these
visions has been an arduous effort, but we are now entering a new epoch
as LGS systems are commissioned on the largest ground-based telescopes.
The scientific promise of near diffraction-limited imaging and
spectroscopy from the ground over most of the sky is finally being
realized.

At this key juncture, the purpose of this paper is to briefly review
published LGS science to date; to provide a snapshot of the science that
is being done with LGS AO; and to look ahead to a future path where LGS
AO is a ubiquitous tool for observational astronomy.

\begin{table}[t]
\vskip -0.1in
\caption{Refereed science papers from LGS AO as of May 2006, listed by
  publication date.  
Some titles have been truncated and some title words abbreviated
(``LGS'' and ``AO''). ``Field'' gives the area of study: ``SS'' = solar
system, ``Gal'' = galactic, ``Xgal'' = extragalactic.  ``$N_{obj}$''
indicates the number of science targets/fields observed with LGS.}
\label{tab:bib}
\small
\begin{center}       
\vskip -0.2in
\begin{tabular}{c|p{1.2in}|p{2.75in}|p{0.33in}|c|c|c} 
\hline
\hline
\rule[-1ex]{0pt}{3.5ex}  
\# & Authors, Journal & Title & Facilty & Field & $\lambda\lambda$ & $N_{obj}$  \\
\hline
\rule[-1ex]{0pt}{3.5ex}
 1 & McCullough \etal\ 1995, ApJ   & Photoevaporating Stellar Envelopes Observed with Rayleigh Beacon Adaptive Optics & SOR & Gal & H$\alpha$ & 1 \\ \hline
\rule[-1ex]{0pt}{3.5ex}
 2 & Christou \etal\ 1995, ApJ     & Rayleigh Beacon AO Imaging of ADS 9731: Measurements of the Isoplanatic Field of View & SOR & Gal & $IJH$ & 1 \\ \hline
\rule[-1ex]{0pt}{3.5ex}
 3 & Drummond \etal\ 1995, ApJ     & Full AO Images of ADS 9731 and $\mu$ Cassiopeiae: Orbits and Masses & SOR & Gal & $IJH$ & 1 \\ \hline
\rule[-1ex]{0pt}{3.5ex}
 4 & ten Brummelaar \etal\ 1996, AJ & Differential Binary Star Photometry Using the AO System at Starfire Optical Range & SOR & Gal & $ri$ & 10 \\ \hline
\rule[-1ex]{0pt}{3.5ex}
 5 & Glenar \etal\ 1997, PASP      & Multispectral Imagery of Jupiter and Saturn Using AO and Acousto-Optic Tuning & SOR & SS & 0.7--1.0 \micron & 2 \\ \hline
\rule[-1ex]{0pt}{3.5ex}
 6 & Koresko \etal\ 1997, ApJ      & A Multiresolution Infrared Imaging Study of LkH$\alpha$ 198 & SOR & Gal & $H$ & 1 \\ \hline
\rule[-1ex]{0pt}{3.5ex}
 7 & Drummond \etal\ 1998, Icarus  & Full AO Images of Asteroids Ceres and Vesta: Rotational Poles \& Triaxial Ellipsoid Dimensions & SOR & SS & $i$ & 2 \\ \hline
\rule[-1ex]{0pt}{3.5ex}
 8 & Barnaby \etal\ 2000, AJ       & Measurements of Binary Stars with the Starfire Optical Range AO System & SOR & Gal & $ri$ & 1 \\ \hline
\rule[-1ex]{0pt}{3.5ex}
 9 & Hackenberg \etal\ 2000, AA    & Near-Infrared AO Observations of Galaxy Clusters: Abell 262 at z=0.0157 ... & Calar Alto & Xgal & $K$ & 1 \\ \hline 
\rule[-1ex]{0pt}{3.5ex}
10 & Perrin \etal\ 2004, Science   & Laser Guide Star AO Imaging Polarimetry of Herbig Ae/Be Stars & Lick & Gal & $JHK_S$ & 2 \\ \hline
\rule[-1ex]{0pt}{3.5ex}
11 & Melbourne \etal\ 2005, ApJL   & Merging Galaxies in GOODS-S: First Extragalactic Results from Keck Laser AO & Keck & Xgal & $K^\prime$ & 1 \\ \hline
\rule[-1ex]{0pt}{3.5ex}
12 & Gal-Yam \etal\ 2005, ApJL     & A High Angular Resolution Search for the Progenitor of the Type Ic  Supernova 2004gt & Keck & Xgal & $K_S$ & 1 \\ \hline
\rule[-1ex]{0pt}{3.5ex}
13 & Brown \etal\ 2005, ApJL       & Keck LGS AO Discovery and Characterization of a Satellite to Large Kuiper Belt Object 2003 EL61 & Keck & SS & $K^\prime$ & 1 \\ \hline
\rule[-1ex]{0pt}{3.5ex}
14 & Muno \etal\ 2005, ApJ         & A Remarkable Low-Mass X-ray Binary within 0.1 pc of the Galactic Center & Keck & Gal & $K^{\prime}L^{\prime}$ & 1 \\ \hline
\rule[-1ex]{0pt}{3.5ex}
15 & Liu \etal\ 2005, ApJ          & Kelu-1 is a Binary L Dwarf: First Brown Dwarf Science from Laser Guide Star AO & Keck & Gal & $JHK^\prime$ & 1 \\ \hline
\rule[-1ex]{0pt}{3.5ex}
16 & Cohen \etal\ 2005, ApJL       & To Be or Not to Be: Very Young Globular Clusters in M31 & Keck & Xgal & $K^\prime$ & 6 \\ \hline
\rule[-1ex]{0pt}{3.5ex}
17 & Ghez \etal\ 2005, ApJ         & The First LGS AO Observations of the Galactic Center: Sgr A*'s Infrared Color ... & Keck & Gal & $K^{\prime}L^{\prime}$ & 1 \\ \hline
\rule[-1ex]{0pt}{3.5ex}
18 & Marchis \etal\ 2006, Nature   & A Low Density of 0.8 g cm$^{-3}$ for the Trojan Binary Asteroid 617 Patroclus & Keck & SS & $HK^\prime$ & 1\\ \hline
\rule[-1ex]{0pt}{3.5ex}
19 & Brown \etal\ 2006, ApJ        & Satellites of the Largest Kuiper Belt Objects & Keck & SS & $K^\prime$ & 4 \\ \hline
\rule[-1ex]{0pt}{3.5ex}
20 & Krabbe \etal\ 2006, ApJL      & Diffraction Limited Imaging Spectroscopy of the SgrA* Region Using OSIRIS & Keck & Gal & 2.0--2.4 \micron & 1 \\ \hline
\rule[-1ex]{0pt}{3.5ex}
21 & Gelino \etal\ 2006, PASP      & Evidence of Orbital Motion in Binary Brown Dwarf Kelu-1AB & Keck & Gal & $HK^\prime$ & 1\\ \hline
\rule[-1ex]{0pt}{3.5ex}
22 & Liu \etal\ 2006, ApJ          & SDSS J1534+1615AB: A Novel T Dwarf Binary Found with Keck LGS AO and the Role of Binarity in the L/T Transition & Keck & Gal & $JHK^\prime$ & 1 \\ \hline
\rule[-1ex]{0pt}{3.5ex}
23 & Sheehy \etal\ 2006, ApJ       & Constraining the AO PSF in Crowded Fields: Measuring Photometric Aperture Corrections & Keck & Xgal & $H$ & 1 \\ \hline
\end{tabular}
\end{center}
\end{table} 

\section{ASTRONOMICAL SCIENCE FROM LASER GUIDE STARS}

To date, four telecopes have produced astronomical science with LGS: the
1.5-meter Starfire Optical Range (SOR) Telescope in New
Mexico\cite{1994OSAJ...11..310F}; the 3-meter Shane Telescope at Lick
Observatory in California\cite{max97}; the 3.5-meter German-Spanish
Astronomical Centre Telescope in Spain\cite{2000ExA....10....1E}; and
the 10-meter Keck~II Telescope in Hawaii\cite{2006PASP..118..297W}.  The
Starfire system used a Rayleigh-backscattered LGS, with the other
systems using sodium laser guide stars.

The first refereed science paper from LGS AO was a study of the Orion
Nebula region using the 1.5-meter telescope at Starfire Optical
Range.\cite{1995ApJ...438..394M} As the authors pointed out: "it is a
truism that if a new telescope or a new instrument can be pointed at
[the Orion Nebula], it will be."  They obtained H$\alpha$ (6563~\AA)
emission-line imaging of the photoevaporating circumstellar envelopes
around the young stars.  The poor natural seeing of the Starfire site
and the low elevation of the Orion Nebula as viewed from New Mexico
limited the AO system to only modest image correction, with image FWHMs
of 0.4\arcsec.  Nevertheless, the LGS-sharpened images enabled detection
of the cometary structure of these systems, demonstrating that good
science opportunities can be harvested even when AO systems do not reach
the diffraction limit.\footnote{Indeed, the McCullough \etal\ (1995)
  work remains the most highly cited of all LGS papers to date, with 40
  citations in ADS at the time this review was written, more than double
  the next most cited LGS paper.}  The resolved circumstellar
morphologies combined with quantitative modeling of the H$\alpha$ and
radio continuum fluxes supported the interpretation of these sources as
undergoing photoevaporation due to the central massive star
$\Theta^1$C~Ori.

It is interesting to note that this first LGS science paper contained
many of the characteristics and concerns that persist today for LGS
astronomical science: (1) uncertainty in the PSF quality (see their
Figure 3 for an analysis of anisoplanatism in their images); (2) using
LGS photometry only for relative photometry of images, with absolute
photometry derived from a seeing-limited dataset; (3) the need for
modeling of the PSF (which was handled using multiple two-dimensional
Gaussians), and (4) comparison with {\sl Hubble Space Telescope} (\HST)
imaging results\cite{1993ApJ...410..696O}.

\begin{figure}[t]
  \begin{center}
    \begin{tabular}{c}
      \includegraphics[height=8cm]{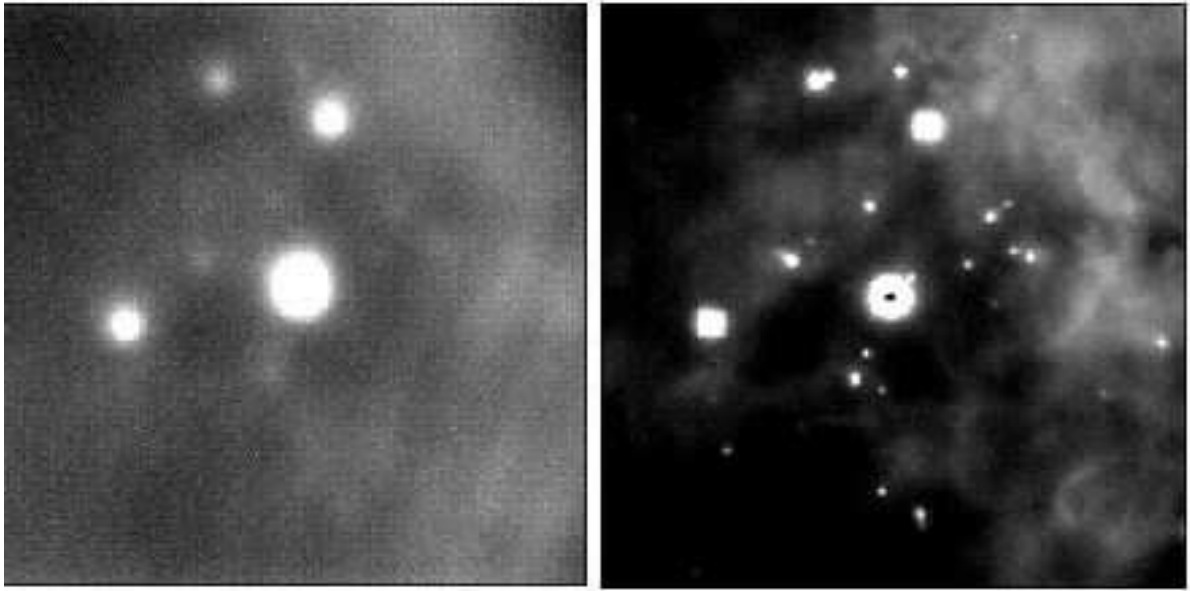}
    \end{tabular}
  \end{center}
  \vskip -4ex
  \caption[fig:mccullough] 
          { \label{fig:mccullough} First astronomical science from LGS
            AO, optical (0.65~\micron) imaging of the central
            40\arcsec\ of the Orion Nebula.\cite{1995ApJ...438..394M}
            The image of the left was obtained without AO and the one of
            the right with LGS AO.  This figure was adapted from
            material on the Starfire Optical Range web site; see
            Ref.~\citenum{1995ApJ...438..394M} for a higher quality
            reproduction.}
\end{figure} 

Since the first paper, LGS science productivity has been relatively
modest.  As a point of comparison: as of July 2002, after about a decade
of science operation, AO systems had produced 144 refereed science
papers, nearly entirely derived from natural guide star (NGS) AO
systems.\cite{2003SPIE.4834...84C} A non-systematic search of the NASA
Astrophysical Data System (ADS) abstract database indicates that the
total number of AO science papers has about doubled since then, again
dominated by NGS observations.

\begin{figure}[t]
  \begin{center}
    \begin{tabular}{c}
      \includegraphics[width=3.9in,angle=90]{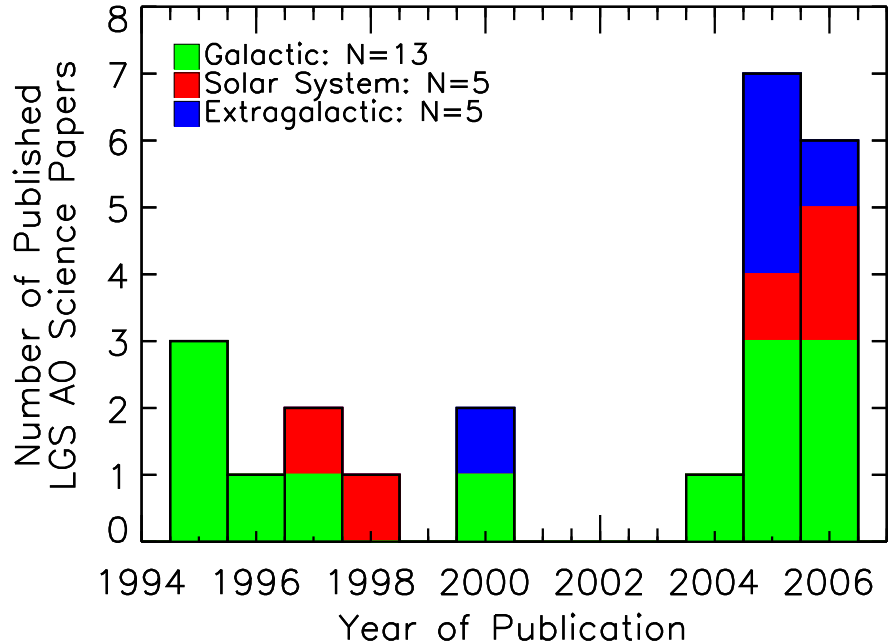}
    \end{tabular}
  \end{center}
  \vskip -0.35in
  \caption[fig:pubs] 
          { \label{fig:pubs} A histogram of all refereed LGS AO science
            papers published as of May 2006.  The colors represent
            different science areas.  The large spike in publications in
            2005--2006 comes from the Keck LGS AO system.}
\end{figure} 

In comparison, LGS systems have produced 23 refereed science
publications as of May 2006, again based on NASA ADS.  The papers are
listed in Table~1 and shown in Figure~1.  This compilation includes only
papers focused on astronomical science; AO instrumentation or "first
light" papers are not included.  The modest number of LGS AO papers
compared to NGS AO is no doubt due to the vastly greater complexity and
cost of LGS systems.  
Nevertheless, the sustained drive over many years for developing LGS AO
points to the wide-spread appreciation for the science potential of this
technology.  The papers from 1995--2004, which were produced by the
Starfire, Calar Alto, and Lick systems, can be categorized as:
\begin{itemize}
\item Binary field stars: 4 papers on resolved photometry and/or
  astrometry \cite{1995ApJ...450..369C, 1996AJ....112.1180B,
    1998Icar..132...80D, 2000AJ....119..378B}
\item Young stars: 3 papers on circumstellar material around pre-main
  sequence stars\cite{1995ApJ...438..394M, 1997ApJ...485..213K,
    2004Sci...303.1345P}
\item Planetary astronomy: 2 papers on resolved imaging of asteroids or
  giant planets\cite{1997PASP..109..326G, 1998Icar..132...80D}
\item Extragalactic science: 1 paper on the nuclear morphology of a
  barred spiral galaxy\cite{2000A&A...363...41H}
\end{itemize}
A number of these papers used post-processing techniques, primarily
deconvolution, to handle the spatially complex and time-variable PSF.
Also, little extragalactic science was done; a significant limiting
factor for LGS AO systems on $\sim$2--3-meter telescopes is the need for
relatively bright tiptilt stars, meaning very modest sky coverage.

The LGS AO system on the Keck~II Telescope has been performing
shared-risk science for about the past 1.5~years (since November~2004)
and has produced all the LGS science papers in 2005--2006 so far.  With
its 10-meter aperture, Keck LGS can provide angular resolution far
exceeding \HST\ at IR wavelengths.  In addition, the large aperture
means the tiptilt star sensitivity ($R\lesssim19$) is sufficient to open
most of the sky to LGS observations (see next section), which is a great
boon for extragalactic science applications.  There has been a
substantial wave of initial science publications from Keck, which can be
categorized into:
\begin{itemize}
\item Extragalactic science: 4 papers\cite{2005ApJ...625L..27M,
  2005ApJ...630L..29G, 2005ApJ...634L..45C, 2006astro.ph..4551S}
\item Binary brown dwarfs: 3 papers\cite{2005astro.ph..8082L,
  2006PASP..118..611G, 2006astro.ph..5037L}
\item Galactic Center: 3 papers\cite{2005ApJ...633..228M,
  2005ApJ...635.1087G, 2006ApJ...642L.145K}
\item Small bodies in the solar system: 3
  papers \cite{2005ApJ...632L..45B, 2006Natur.439..565M,
    2006ApJ...639L..43B}
\end{itemize}
The science breadth and impact of Keck LGS AO has been significant,
including discovery of moons around small bodies in the solar system and
the resulting mass measurements\cite{2005ApJ...632L..45B,
  2006ApJ...639L..43B}, color and variability measurements for Sgr~A$^*$
in the Galactic Center\cite{2005ApJ...635.1087G, 2006ApJ...642L.145K},
studying ultracool atmospheres with a new kind of binary brown
dwarf\cite{2006astro.ph..5037L}, and decomposing the stellar components
of $z\approx0.5$ galaxies\cite{2005ApJ...625L..27M}.  As a point of
reference, almost none of the initial Keck science results could have
been accomplished with the previous generation of LGS AO systems.

\clearpage
\section{CURRENT KECK LGS PERFORMANCE:\\A Near-IR Imaging Survey of
  Nearby Brown Dwarfs} 

To illustrate what is currently possibly with LGS, we now discuss the
on-sky performance of the Keck LGS system.  For about the past year, my
collaborators and I have been conducting a high angular resolution
near-IR imaging survey of nearby brown dwarfs with the Keck LGS system
(\eg, Figure~3).  Our goals are (1) to assess the binary frequency of
ultracool dwarfs; (2) to test atmospheric models with these coeval
systems (\eg, as associated with the abrupt spectral transition from the
L~dwarfs to the T~dwarfs); (3) to search for exceptionally
low-temperature companions; and (4) to find and monitor substellar
binaries suitable for dynamical mass determinations.

LGS AO represents a major instrumental advance for this science area.
Brown dwarfs are too optically faint for natural guide star AO.  Also,
most known brown dwarf binaries have separations of
$\lesssim$0.3\arcsec,\cite{2006astro.ph..2122B} hence the need for high
angular resolution imaging to find and characterize them via resolved
photometry and spectroscopy.  Our observations have achieved
3--4$\times$ the angular resolution at $K$-band (2.2~\micron) compared
to \HST\ and thus are more sensitive to close companions.  In addition,
the ability of Keck LGS AO to find tighter binaries means that systems
with much shorter orbital periods than the current sample can be found
and expeditously monitored.

\begin{figure}[t]
  \begin{center}
    \begin{tabular}{c}
      \includegraphics[width=2.25in]{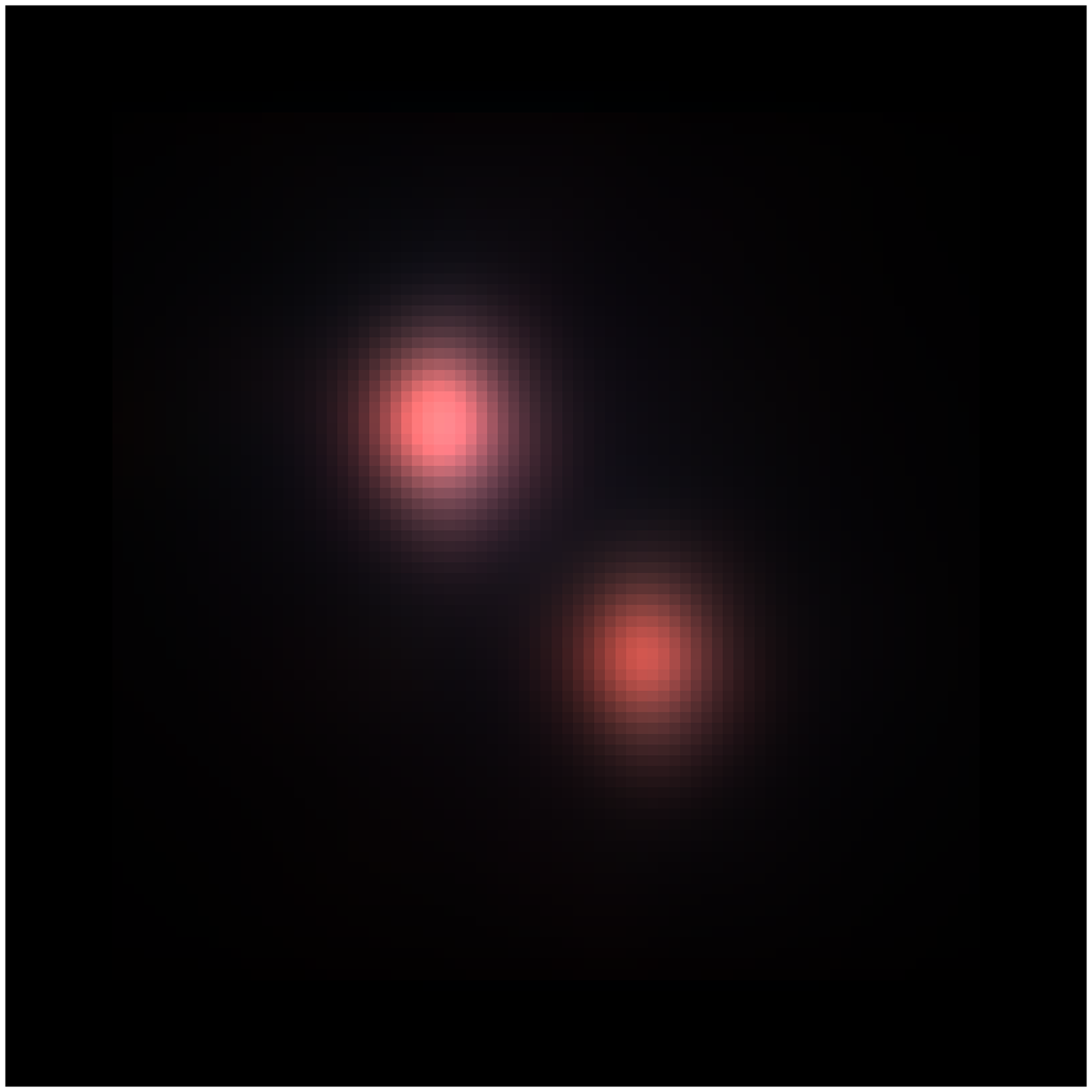}
      \includegraphics[width=2.25in]{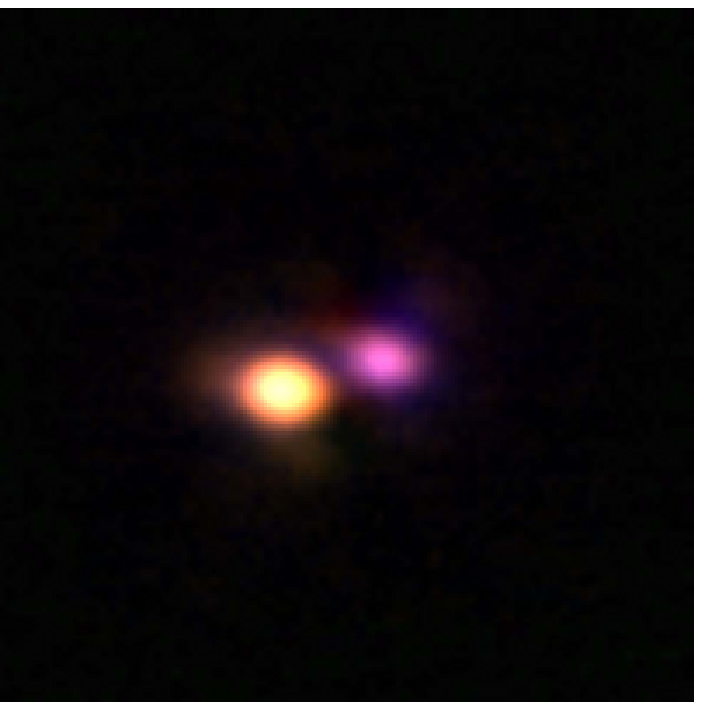}
    \end{tabular}
  \end{center}
  \vskip -3ex
  \caption[fig:binaries] 
          { \label{fig:binaries} Two of the brown dwarf binaries
            discovered with Keck LGS AO, the L2+L4 binary Kelu-1AB ({\em
              left}) and the T1.5+T5.5 binary SDSS 1534+1615AB ({\em
              right}).\cite{2005astro.ph..8082L, 2006astro.ph..5037L}
            The binaries have separations of 0.29\arcsec\ and
            0.11\arcsec, respectively.  These color images were made
            from $JHK^{\prime}$-band data.  The noticeably different
            colors of the two components of SDSS~1534+1615AB point to
            very different photospheric dust content, despite their
            similar masses and temperatures.\cite{2006astro.ph..5037L}}
\end{figure} 

Our brown dwarf imaging survey provides an excellent dataset for
assessing typical Keck LGS performance in the case of off-axis
observations, namely the situation where the LGS is pointed to the
science target but tiptilt sensing and correction are derived from an
adjacent field star.  Brown dwarfs are far too optically faint to serve
as their own tiptilt references and hence the need for a nearby star --
this is the same observing situation as expected for many extragalactic
LGS applications and thus provides a good reference point.  For Keck,
the tiptilt star must be within 60\arcsec\ of the science target -- in
practice, we find that this results in a sky coverage fraction of about
2/3 for an estimated $K$-band Strehl ratio of $\gtrsim$0.2.  Since
wide-field brown dwarf searches encompass most of the sky (except for
avoidance of the galactic plane), this 2/3 sky coverage estimate is a
fair representation of the fraction of any set of generic targets that
can be imaged with LGS.

Figure~4 summarizes the quality of Keck LGS observations to date, based
on multiple observing runs over the past year.  No bad data have been
censored, so a mix of seeing conditions and technical performance (e.g.,
LGS projected power and sodium light return flux) are represented.  (See
also paper 6272-01 by LeMignant \etal\ in this Proceedings.)  The median
$K$-band image FWHM for our survey is 0.069\arcsec\ with a best value of
0.049\arcsec.  The median Strehl is 0.18 with a best value of 0.43.
Good performance is achieved for tiptilt stars approaching
$R\approx18$~mag, with best performance for $R\lesssim17$~mag.  We have
successfully used tiptilt stars as widely separated as 60\arcsec\ from
the science target, which represents the outer range of the tiptilt
stage field of regard.

\begin{figure}[h]
  \begin{center}
    \begin{tabular}{c}
      \hskip -0.3in
      \includegraphics[width=7in]{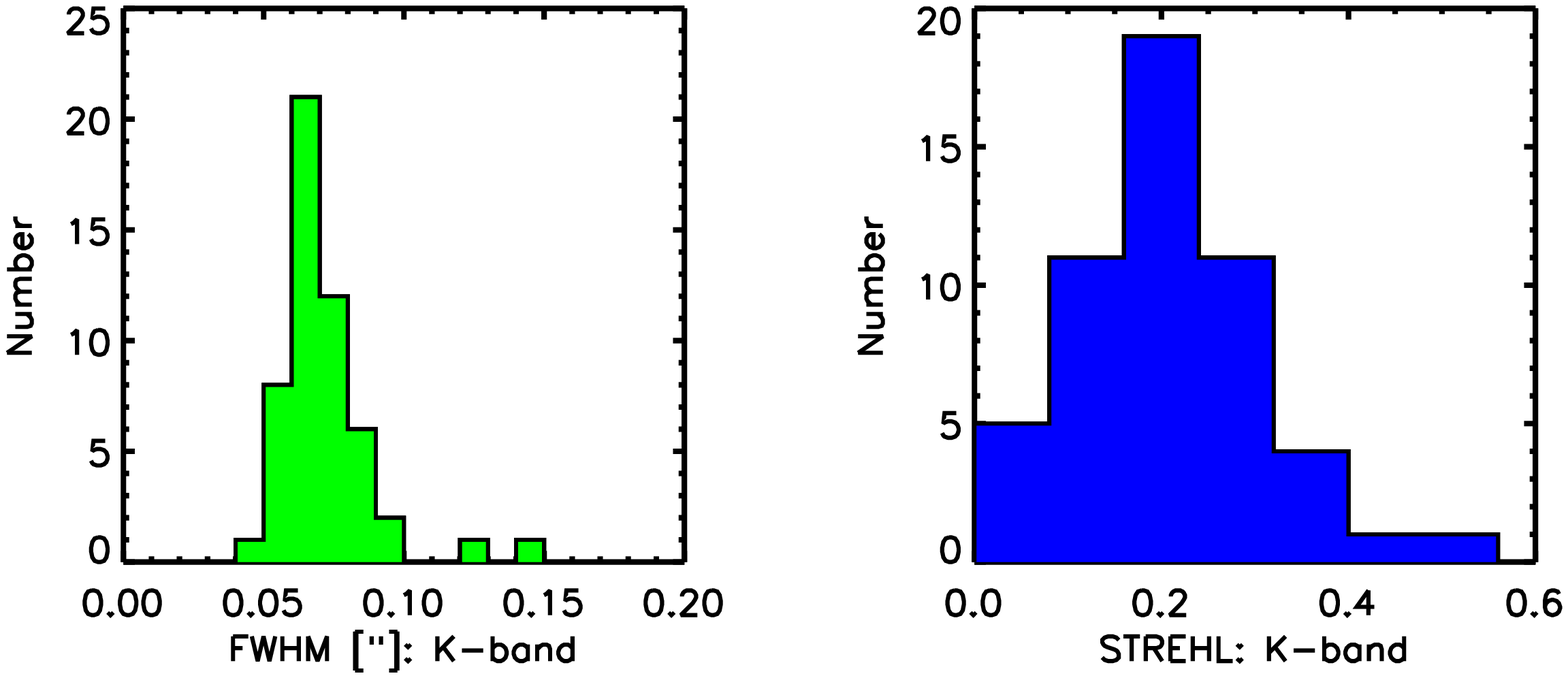}\\
      \hskip -0.3in
      \includegraphics[width=7in]{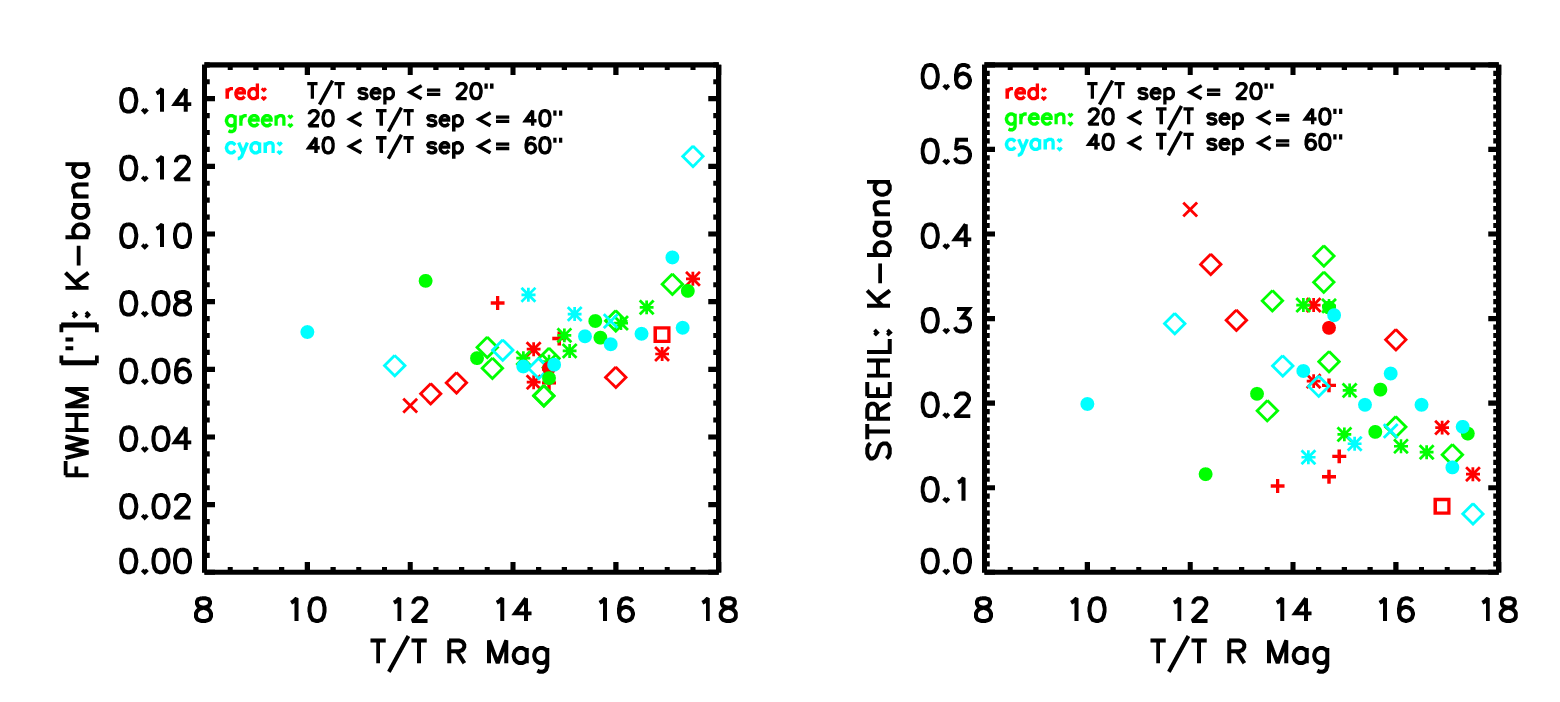}
    \end{tabular}
  \end{center}
  \vskip -0.3in
  \caption[fig:kecklgs] 
          { \label{fig:kecklgs} Summary of Keck LGS AO $K$-band
            (2.2~\micron) performance, based on our near-IR imaging
            survey of brown dwarfs.  Each data point represents the
            average for a set of images of a given object.  No bad data
            have been censored, so a mix of seeing conditions, target
            airmasses, and technical performance are represented.  {\bf
              Top panels:} Histogram of FWHM and Strehl ratios.  The
            median FWHM is 0.069\arcsec\ with a best value of
            0.049\arcsec.  The median Strehl is 0.18 with a best value
            of 0.43.  {\bf Bottom panels:} Image quality are a function
            of tiptilt star $R$-band magnitude, as listed in the
            USNO-B1.0 catalog.\cite{2003AJ....125..984M} Different
            colors represent the angular separations of the tiptilt
            stars from the science targets, and different symbols
            represent different observing runs.}
\end{figure}

\clearpage
\section{LASER GUIDE STARS VERSUS HUBBLE SPACE TELESCOPE}

The advent of LGS AO on 8--10 meter class telescopes naturally leads to
the question of the relative roles of high angular resolution astronomy
from the ground and from space.  While implementation of LGS AO systems
is ongoing, 
there are already several clear advantages of LGS compared to
\HST:

\begin{itemize}
\item {\em Superior angular resolution in the near-IR:} Keck LGS
  produces images as sharp as 0.05\arcsec\ FWHM at $H$ and $K$-bands,
  about 3--4$\times$ better than HST.  In general, LGS AO excels at
  studying close point sources, such as tight
  binaries\cite{2005ApJ...632L..45B,2005astro.ph..8082L,
    2006Natur.439..565M, 2006ApJ...639L..43B, 2006PASP..118..611G,
    2006astro.ph..5037L} and dense star
  clusters\cite{2005ApJ...635.1087G, 2006astro.ph..4551S}.  LGS is
  likely to become a popular platform for these kinds of near-IR
  observations.  In addition, LGS in the near-IR offers comparable
  angular resolution to \HST\ at optical wavelengths, allowing for
  well-matched imaging studies over a wide-range of
  wavelengths\cite{2005ApJ...625L..27M, 2005ApJ...630L..29G}.

\item {\em Ability to obtain thermal IR imaging (3--4~\micron) and
  spatially resolved IR (1--5~\micron)
  spectroscopy\cite{2005ApJ...633..228M, 2005ApJ...635.1087G,
    2006ApJ...642L.145K}:} Neither of these capabilities is available
  from \HST.

\item {\em Efficient survey-style observing:} LGS AO can observe many
  ($>$20) targets in one night, whereas \HST\ is limited to about
  one~target per orbit.  Thus, LGS AO can carry out high angular
  resolution surveys of many objects, which would otherwise require a
  prohibitive number of \HST\ observing time.  Figure~5 shows the target
  acquisition times experienced by our Keck LGS survey, including
  telescope slew, LGS propagation, and AO performance optimization.  The
  average setup time is 9~minutes, with times as short as 5~minutes and
  a long tail in the distribution due to occasional significant
  technical problems.

\begin{figure}
  \vskip -0.2in
  \begin{center}
    \begin{tabular}{c}
      \hskip -0.5in
      \includegraphics[width=3.5in,angle=90]{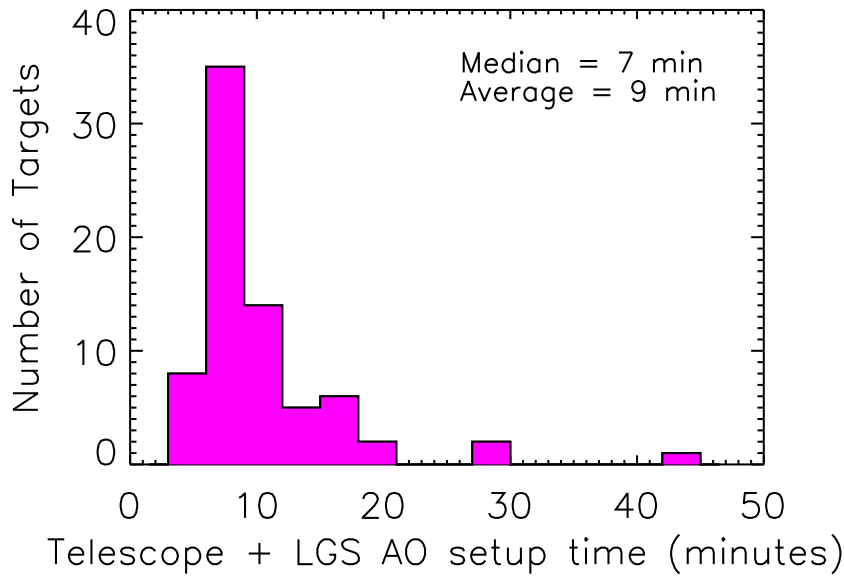}
    \end{tabular}
  \end{center}
  \vskip -6ex
  \caption[fig:acquisition] 
          { \label{fig:acquisition} Target acquisition times for Keck
            LGS AO observing, from the first year of our brown dwarf
            imaging survey.  These include telescope slew, field
            acquisition, LGS propagation, AO system optimization, and
            technical difficulties, but does not include the
            $\approx$30~min needed at the start of each night for
            initial LGS AO setup.}
\end{figure}

\item {\em Long-term availability:} LGS AO offers an enduring platform
  for multi-epoch high angular resolution science, \eg, monitoring of
  time-variable phenomena or orbital motion.\footnote{An interesting
    illustration of this is provided by the case of Kelu-1.  One of the
    first free-floating brown dwarfs discovered in the solar
    neighborhood\cite{1997ApJ...491L.107R}, this nearby L~dwarf had long
    been noted to be overluminous compared to other similar
    objects\cite{1999AJ....118.2466M, leg01}.  \HST\ imaging in 1998
    failed to identify any close companion,\cite{1998ApJ...509L.113M}
    leading to the intriguing possibility that Kelu-1's high luminosity
    was due to a very young age ($\sim$10~Myr); the implied mass would
    have been only $\approx$12~\Mjup\cite{gol04}.  Keck LGS imaging in
    2005 discovered Kelu-1 to be a 0.29\arcsec\ binary and showed that
    binarity explains its many anomalous properties (high luminosity,
    very red color, high inferred effective temperature, and low lithium
    absorption)\cite{2005astro.ph..8082L}.  The projected separation of
    the binary was too small to be easily resolved at the time of the
    original \HST\ imaging,\cite{2006PASP..118..611G} but subsequent
    orbital motion over 7~years enabled the binary to be found with Keck
    LGS AO.}  In contrast, the long-term future is \HST\ is uncertain.

\item {\em Novel instrumentation tailored for LGS AO:} Examples here include
  Keck's near-IR integral field spectrograph
  OSIRIS\cite{2003SPIE.4841.1600L} and the dual-channel imaging
  polarimeter at Lick Observatory\cite{2004Sci...303.1345P}.
\end{itemize}

\noindent And the current disadvantages of LGS compared to \HST\ include:

\begin{itemize}
\item {\em Limited to the IR wavelengths:} LGS AO offers best performance at
  $H$-band (1.6~\micron) and longer wavelengths. High angular resolution
  imaging at optical wavelengths remains the domain of \HST.

\item {\em Complex point spread function (PSF):} The LGS AO PSF is
  irregular in appearance and varies in time. Therefore, in contrast to
  imaging of point sources, LGS studies of extended sources (``fuzzy
  blob science'') is especially challenged by the uncertainties in the
  PSF at the time of observation.  This is also true for studies of very
  faint point sources next to much brighter point sources, where image
  contrast is the key requirement rather than pure angular resolution.
  In these cases, the exact dividing line between \HST\ and LGS AO will
  depend on the particular science goals.

\item {\em Limited field of view:} Good LGS correction is restricted to
  the isoplanatic angle, $\approx$30-40\arcsec\ at $K$-band, and the PSF
  is spatially variable across this field.

\item {\em Need for a tiptilt star:} Only about 2/3 of the sky is
  accessible to LGS AO, due to the need for the tiptilt star close to
  the science target --- specific objects of interest may only be
  observable by \HST.  This sky coverage fraction is an approximate
  average over the entire sky; LGS coverage will be worse near the
  galactic poles (\eg, the Hubble Deep Field) and better close to the
  galactic plane.

\item {\em Heterogenous data:} LGS AO performance can vary greatly, as
  it depends on seeing conditions, laser+AO performance, and tiptilt
  star properties (Figure~4).  \HST\ offers much more predictable
  performance.
\end{itemize}

\section{CONCLUDING THOUGHTS:\\EXPECTATIONS AND WISHES FOR THE
  (NOT-TOO-DISTANT) FUTURE} 

More than two decades since its conception, LGS AO is now entering a new
phase in its growth, opening a new era in high angular resolution
science.  The first decade of astronomical science was marked by modest
science productivity, as these LGS AO systems were commissioned and
optimized ($\approx$1~science paper/year).  Benefiting from the fruits
of this effort, the LGS AO system on the Keck~II 10-meter telescope has
had a highly successful first 1.5 years of science
($\approx$1~paper/month).  Several LGS systems are planned to come
online in the next two years, including those at the Gemini-North
(2006), VLT (2006), Palomar 5-meter (2006), MMT (2006), Subaru (2007),
Gemini-South (2007), and the Keck~I (2008) telescopes.  Given the size
and quality of the science communities associated with these new AO
systems, we can look forward to even more significant growth in LGS
science soon ($\approx$1~paper/week?).

We conclude with a non-comprehensive wishlist for future LGS
developments, in order to highlight ongoing efforts and to review some
outstanding challenges:

\begin{itemize}

\item {\em Field of view:} Single LGS systems produce a corrected field
  of view the size of the isoplanatic angle.  In fact, nearly all LGS
  science thus far has been restricted to objects that span
  $\lesssim$10\arcsec\ across (and most typically
  $\lesssim2$\arcsec). The upcoming Gemini-South multi-conjugate AO
  system is notable, as it will be the first LGS AO system to correct
  much larger fields of view ($\approx$1--2\arcmin).

\item {\em Observing efficiency:} Nighttime LGS AO operations currently
  require many more personnel compared to regular seeing-limited
  observations.  (See also paper 6270-12 by LeMignant \etal\ in this
  Proceedings.)  Greater automation will reduce this burden on telescope
  staff and also should lead to greater efficiency, as observing
  procedures are streamlined.  For ``survey-style'' science programs, in
  some sense the science return is proportional to the number of targets
  observed; therefore more efficient observing is a significant benefit.

\item {\em Robust real-time performance predictions:} LGS image quality
  varies depending on nightly weather conditions, laser performance, AO
  performance, sodium layer density and structure, and tiptilt star
  properties.  To maximize the science return, it would be desireable to
  be able to robustly predict LGS performance for any given target on
  any given night.  This capability would allow observations to be
  tailored to achieve the desired science goal.  Queue-scheduled
  observing is a key element here, but also better real-time
  understanding of seeing conditions, LGS performance and the conditions
  in the sodium layer are need.


\item {\em Near-IR tiptilt sensors:} While much of the sky is available
  for LGS, the most obscured regions (\eg, star-forming regions) are not
  due to the lack of optically visible tiptilt stars.  Tiptilt sensors
  working at near-IR wavelengths would open the door to studying the
  youngest stages of star and planet formation.

\item {\em High quality catalogs for tiptilt stars:} Our brown dwarf
  imaging survey finds about 1 in 10 tiptilt stars with
  $R\lesssim18$~mag from the USNO-B1.0 catalog\cite{2003AJ....125..984M}
  are unsuitable, either because they turn out to be faint galaxies or
  they turn out to be much fainter than the reported magnitudes.  The
  Pan-STARRS project\cite{2002SPIE.4836..154K} will provide precise,
  multi-band photometry over the entire sky visible from Hawaii, with
  the initial PS-1 telescope beginning operations later this year.
  Combined with the SkyMapper Telescope\cite{2005AAS...206.1509S} in the
  southern hemisphere, high quality all-sky catalogs should be available
  in a few years for robust selection of tiptilt stars.

\item {\em Improved PSF stability and characterization:} The
  time-variability of the LGS PSF will remain a significant concern for
  the foreseeable future.  Post-processing software techniques can
  provide some immediate assistance, \eg, deconvolution and/or PSF
  modeling techniques.\cite{1993ApJ...415..862J, 2000A&AS..147..335D,
    2006astro.ph..4551S} Development of algorithms to use AO telemetry
  data to estimate the real-time PSF would be a valuable capability for
  LGS.\cite{2005PASP..117..847S} On a somewhat longer timescale,
  instruments tailored to handle the challenges of LGS AO imaging, such
  as dual-channel imaging systems, can circumvent this problem for some
  types of science programs.  Finally, next-generation LGS AO systems
  should produce higher Strehl imaging, leading to more stable and
  well-behaved PSFs.

\end{itemize}

While the technology is far from mature, LGS AO is entering a phase of
rapid growth.  There is little doubt that it will quickly become a key
capability for a very broad range of astrophysics, spanning the nearest
solar system bodies to the highest redshift galaxies and the entire
universe in-between.

\acknowledgments     

It is pleasure to thank the many, many, many people who have imagined,
advocated, designed, constructed, tested, commissioned, supported,
and/or persevered to make astronomy with laser guide stars a reality.
We also gratefully acknowledge the teams at Laurence Livermore National
Laboratory, Lick Observatory, and Keck Observatory for stoking our
enthusiasm for LGS AO.  We thank Dagny Looper for the Keck LGS
performance analysis, Peter Wizinowich for providing the Keck LGS
bibliography, and several colleagues for proofreading the contents of
Table~1.  MCL's research presented herein is partially supported from
NSF grant AST-0507833 and an Alfred P. Sloan Research Fellowship.  We
wish to recognize and acknowledge the very significant cultural role and
reverence that the summit of Mauna Kea has always had within the
indigenous Hawaiian community.  We are most fortunate to have the
opportunity to conduct observations from this mountain.



\end{document}